\documentclass[journal ]{IEEEtran}
\usepackage{amsmath}
\usepackage{graphicx}
\usepackage{subfigure}
\usepackage{multirow}
\usepackage{enumerate}
\usepackage{setspace}
\usepackage{booktabs}
\usepackage{cite}
\usepackage{caption}
\usepackage{float}
\usepackage{color}
\usepackage{amssymb}
\usepackage{makecell}

\ifCLASSINFOpdf

\else

\fi

\begin{document}

\title{Unsupervised Spatial-spectral Network Learning for Hyperspectral Compressive Snapshot Reconstruction}

\author{

\IEEEauthorblockN{Yubao Sun, Ying Yang, Qingshan Liu, \IEEEmembership{Senior Member,~IEEE}, Mohan Kankanhalli, \IEEEmembership{Fellow,~IEEE}}

\thanks{ This work was supported in part by the National Natural Science Foundation of China under Grant U2001211, 61825601 and 61672292. (Corresponding author: Qingshan Liu.)
}
\thanks{ Y. Sun, Y. Yang and Q. Liu are with the Engineering Research Center of Digital Forensics, Ministry of Education, and the Collaborative Innovation Center of Atmospheric Environment and Equipment Technology, Nanjing University of Information Science and Technology, Nanjing 210044, China. E-mail: sunyb@nuist.edu.cn, yingyang@nuist.edu.cn, qsliu@nuist.edu.cn.}
\thanks {M. Kankanhalli is with the School of Computing, National University of Singapore, Singapore. E-mail: mohan@comp.nus.edu.sg.}
}

%
%

\markboth{}
{Shell \MakeLowercase{\textit{et al.}}: Bare Demo of IEEEtran.cls for IEEE Journals}

%



\maketitle

\begin{abstract}
Hyperspectral compressive imaging takes advantage of compressive sensing theory to achieve coded aperture snapshot measurement without temporal scanning, and the entire three-dimensional spatial-spectral data is captured by a two-dimensional projection during a single integration period. Its core issue is how to reconstruct the underlying hyperspectral image using compressive sensing reconstruction algorithms. Due to the diversity in the spectral response characteristics and wavelength range of different spectral imaging devices, previous works are often inadequate to capture complex spectral variations or lack the adaptive capacity to new hyperspectral imagers. In order to address these issues, we propose an unsupervised spatial-spectral network to reconstruct hyperspectral images only from the compressive snapshot measurement. The proposed network acts as a conditional generative model conditioned on the snapshot measurement, and it exploits the spatial-spectral attention module to capture the joint spatial-spectral correlation of hyperspectral images. The network parameters are optimized to make sure that the network output can closely match the given snapshot measurement according to the imaging model, thus the proposed network can adapt to different imaging settings, which can inherently enhance the applicability of the network. Extensive experiments upon multiple datasets demonstrate that our network can achieve better reconstruction results than the state-of-the-art methods.
\end{abstract}


%
\IEEEpeerreviewmaketitle

\section{Introduction}\label{sec:introduction}

\IEEEPARstart{H}{yperspectral} imaging aims at sampling the spectral reflectance of a scene to collect a three-dimensional (3D) dataset consisting of two spatial dimensions $(h,w)$ and one spectral dimension $\lambda$, called a data-cube $(h,w,\lambda )$ \cite{hagen2013review}. Compared with panchromatic images, the sensed spectral signature of each pixel in hyperspectral images (HSIs) covers a broad range of wavelengths and has a high spectral resolution, so it can reveal more properties of objects at the corresponding spatial position in the scene. Hyperspectral imaging has the superiority of achieving more accurate object classification, and has been acted as a useful tool in many applications including environmental remote sensing \cite{borengasser2007hyperspectral}, land cover classification \cite{cao2018hyperspectral}, anomaly detection \cite{xu2015anomaly} and material identification \cite{hagen2013review,ojha2015spectral,attas2003near}.

Many different techniques have been developed for acquiring 3D hyperspectral cubes \cite{cao2016computational,lin2014spatial,baek2017compact,schechner2002generalized,du2009prism}. These imaging systems usually only sample one or two dimensions of the data-cube at a time, and then employ spectral or spatial scanning to complement the remaining dimensions. For instance, spatially scanned hyperspectral imaging systems \cite{mouroulis2000pushbroom} measure the $(h,\lambda )$ slices of the data-cube by a two-dimensional sensor array in push-broom imaging spectrometers, or collect only one point of the data-cube in whisk-broom imaging spectrometers, and spatial scanning is performed to capture the entire data-cube. Spectrally scanned hyperspectral imaging systems, such as fixed or tunable filter spectrometers, sense a single spectral band $(h,w)$ of the data-cube at a time and cover all the spectral bands by scanning along the spectral dimension \cite{Nahum2000tuablefilter}. However, these spectrometers may suffer from motion artifacts during the period of scanning. Furthermore, since the light collection efficiency of the entrance slit in these spectrometers is insufficient, the imaging quality may be degraded.

\begin{figure*}[htbp]
\centering
\includegraphics [scale=0.57]{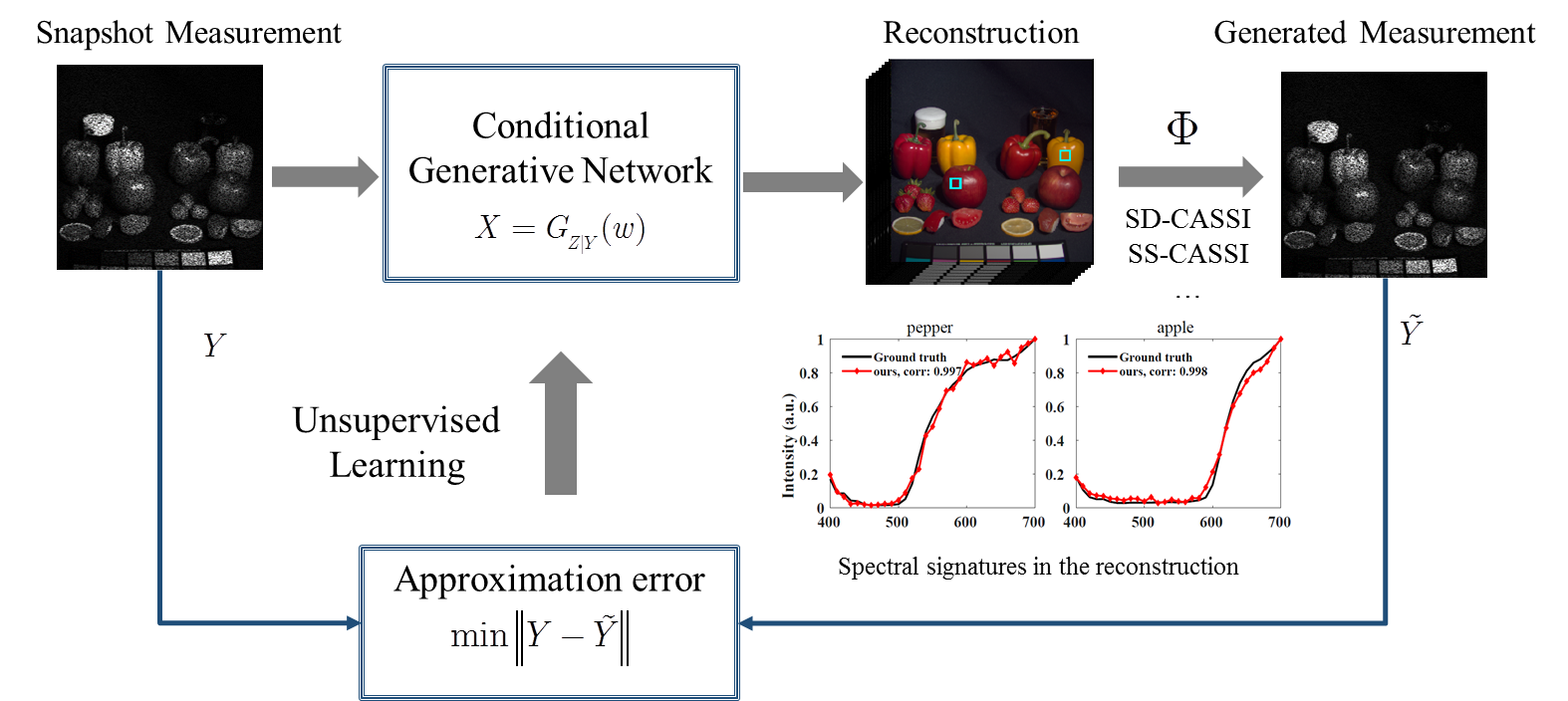}
\caption{The flowchart of the proposed hyperspectral compressive snapshot reconstruction method based on unsupervised network learning. The detailed network architecture of conditional generative network $G$ is shown in Fig. \ref{csnet}.}
\label{framework}
\end{figure*}

Different from the scanning hyperspectral imaging systems mentioned above \cite{porter1987system,schechner2002generalized}, snapshot imaging spectrometers collect both the spectral and spatial information in a single integration period. Therefore, motion artifacts can be avoided in snapshot imaging, and the light collection efficiency can also be significantly improved, enabling the capture of dynamic scenes. Coded aperture snapshot spectral imaging (CASSI) \cite{gehm2007single} is one of the well-known hyperspectral snapshot imaging systems. It takes advantage of the compressed sensing (CS) technology and achieves a two-dimensional (2D) snapshot measurement by random linear projection. Specifically, the snapshot measurement in CASSI is calculated as the linear encoding of the scene through a random code aperture mask. There are mainly two encoding manners \cite{choi2017high}, one is encoding the optical field in the spatial dimension with a single disperser like SD-CASSI \cite{wagadarikar2008single}, and the other is encoding the optical field in both spatial and spectral dimension, like SS-CASSI \cite{lin2014spatial} or DD-CASSI \cite{gehm2007single}. The encoded light field is then integrated by a 2D detector array. An optimization algorithm needs to be used to reconstruct the spectral scenes from the 2D snapshot measurement, which is far fewer than the number of samples required by conventional scanning based spectrometers.

Due to the under-determined observations in CASSI, HSI reconstruction from the snapshot measurement is an ill-posed inverse problem. To deal with this issue, some hand-crafted priors have been designed to represent hyperspectral images, including total variation (TV) \cite{kittle2010multiframe}, $l_1$ sparsity \cite{wagadarikar2008single}, low-rank \cite{martin2014hyca}, and non-local self-similarity. The reconstruction can be obtained by solving these priors regularized optimization problems. However, these prior structures are designed empirically and are therefore insufficient to represent the complicated spectral variation of the real-world scenes. With the powerful learning capabilities of deep networks \cite{lecun2015deep,gatys2016image,jin2017deep,zhang2017beyond}, some works attempted to learn the parameterized network representation of HSIs in a data-driven manner \cite{choi2017high}. However, they all require a large number of hyperspectral images for supervised learning. In practical scenarios, it is expensive to collect enough training data sets for network pre-training. In addition, due to the differences in the spectral response characteristics and spectral wavelength range of different spectral imaging devices, the pre-trained network upon some specific hyperspectral datasets usually cannot be well applicable to other hyperspectral imagers.

In order to cope with these issues, we propose an unsupervised network to learn HSI reconstruction only from the compressive snapshot measurement without pre-training. As shown in Fig. \ref{framework}, the proposed network acts as a conditional generative network for generating the underlying hyperspectral images from random code $Z$ conditioned on the given snapshot measurement $Y$. Different from gray or color images, hyperspectral images present a joint correlation among spatial and spectral dimensions. Therefore, the conditional generative network is equipped with specific modules to capture spatial-spectral correlations, which can effectively reconstruct the spatial-spectral information of HSIs. The network parameters are optimized to generate the optimal reconstruction which can closely match the given snapshot measurement according to the imaging model. We refer to our \textbf{H}yperspectral \textbf{C}ompressive \textbf{S}nap\textbf{S}hot reconstruction \textbf{Net}work as \textbf{HCS$^2$-Net} for short. Our main contributions can be summarized as:
\begin{enumerate}
  \item We propose an unsupervised HCS$^2$-Net for hyperspectral compressive snapshot reconstruction, which learns the reconstruction only from the snapshot measurement without pre-training. In practical scenarios, it can greatly enhance the adaptability and generalization due to the characteristics of unsupervised learning.
  \item The spatial-spectral joint attention module is designed to seize the correlation between the spatial and spectral dimensions of HSIs. This module learns multi-scale 3D attention maps to adaptively weight each entry of the feature map, which is beneficial to improve the reconstruction quality.
  \item The proposed HCS$^2$-Net is evaluated upon multiple simulated and real data sets with both the SD-CASSI and SS-CASSI systems. The quantitative results show that HCS$^2$-Net achieves promising reconstruction results, and outperforms the state-of-the-art methods.
\end{enumerate}

The remainder of this paper is organized as follows. In section II, we review some related works,
especially two kinds of popular works, namely the predefined prior-based and deep network-based methods. Section III introduces the CASSI systems, and section IV describes the proposed HCS$^2$-Net, including network architecture and network learning. We report the experimental results in section V and conclude the paper in section VI.

\section{Related work}
The coding-based snapshot imaging methods rely on the principle of compressed sensing \cite{donoho2006compressed}, and the number of entries in the snapshot measurement (such as SD-CASSI and SS-CASSI measurements) is much smaller than the original HSI size. Therefore, this reconstruction problem is under-determined, and a proper prior representation of HSIs is needed to achieve reliable reconstruction. According to the different priors used, the popular HSI compressive snapshot reconstruction methods can be mainly grouped into two categories: predefined prior-based methods and deep network-based reconstruction methods.

\textbf{Predefined Prior-based HSI reconstruction}. This kind of method seeks the reconstruction by optimizing an objective function consisting of a data fidelity term and a regularization term. The data fidelity term penalizes the mismatch between the unknown HSI and the given measurement according to the imaging observation model, and the regularization term constraints the prior structures of HSIs. Many prior structures have been exploited to represent the HSI, such as the sparsity prior, the total variation and the low rank structure \cite{kittle2010multiframe,wagadarikar2008single,martin2014hyca}.

Many studies are developed within this paradigm. \cite{bioucas2007new} and \cite{yuan2016generalized} both choose the prior of total variation (TV) as a regularization term for each band, and they employ the two-step iterative shrinkage thresholding (TwIST) algorithm and the generalized alternating projection (GAP) for model optimization respectively. \cite{yin2008wavelet} represents the unknown signal in the wavelet and DCT domain and solves the induced sparsity-regularized reconstruction problem by the bregman iterative algorithm. Instead of using a predefined transformation, \cite{xin2015dictionary} learns a dictionary to represent the underlying HSI data-cube. Liu $et$ $al$. \cite{liu2018rank} proposed a method dubbed DeSCI to capture the nonlocal self-similarity of HSIs by minimizing the weighted nuclear norm. Compared with the TV regularization, DeSCI can achieve a better reconstruction performance, but it takes a lot of time to carry out patch search and singular value decomposition. Overall, all these prior structures are hand-crafted based on empirical knowledge, so they lack an adaptive ability to spectral diversities and the non-linearity distribution of hyperspectral data. At the same time, these priors also involve the empirical setting of some parameters.

\textbf{Deep Network-based HSI reconstruction}. In recent years, deep neural networks have been proven to achieve the start-of-the-art results for a variety of image-related tasks including image compressive sensing reconstruction \cite{kulkarni2016reconnet,sun2020learning, yang2020admmnet}. Unlike the predefined prior based algorithms, deep network-based methods attempt to directly learn the image prior by training the network on a large number of data sets, thereby capturing the inherent statistical characteristics of HSIs.

Several studies have exploited deep networks for hyperspectral compressive reconstruction from the CASSI measurement. Xiong $et$ $al.$ \cite{xiong2017hscnn} upsampled the undersampled measurement into the same dimension as the original HSI, and enhanced the reconstruction by learning the incremental residual with a convolutional neural network. Choi $et$ $al.$ \cite{choi2017high} designed a convolutional autoencoder to obtain nonlinear spectral representation of HSIs, and they adopted the learned autoencoder prior and total variation prior as a compound regularization term of the unified variational reconstruction problem, and this reconstruction problem was optimized with the alternating direction method of multipliers to obtain the final reconstruction. Wang $et$ $al.$ \cite{wang2018hyperreconnet} proposed a network named HyperReconNet to learn the reconstruction, in which a spatial network and a spectral network were concatenated to finish the spatial-spectral information prediction. Zhang $et$ $al.$ \cite{zhang2019hyperspectral} built a network composed of multiple dense residual blocks with residual channel attention to learn the reconstruction. In order to better cope with the complexity and variability of HSIs, the external dataset and the internal information of input coded image are jointly used in \cite{zhang2019hyperspectral}. Miao $et$ $al.$ \cite{miao2019lambda} proposed a network named $\lambda$-net to learn the reconstruction mapping by a two-stage generative adversarial network. It used a deeper self-attention U-net in the first stage to obtain an initial reconstruction, and another U-net to improve the reconstruction in the first stage.

Although these deep network-based methods can explore the power of deep feature representation for boosting the reconstruction accuracy, they all require large data sets for network pre-training. At the same time, these pre-trained networks are dedicated to a single observation system. When a new coded aperture mask is used in the imaging system, it is necessary to re-train the reconstruction networks for good reconstruction, so the pre-trained networks have a poor generalization ability to other HSI imaging systems. Unsupervised network learning is an effective method to cope with this issue.
Motivated by the deep network prior \cite{ulyanov2018deep}, some networks, including a convolutional neural network with learned regularization \cite{Veen2020csdip} and a non-locally regularized encoder-decoder network \cite{sun2020nlrcsnet}, are developed for compressed sensing of images without network pre-training. However, the compressive imaging principle of HSIs is different from that of monochromatic images \cite{sun2020nlrcsnet}, which is mainly reflected in the complex spatial-spectral joint structures of HSIs. In this paper, we propose an unsupervised spatial-spectral network for hyperspectral compressive snapshot reconstruction, in which a multi-scale spatial-spectral attention module is designed to capture the complex spatial-spectral correlation. The proposed method reconstructs HSIs from the given snapshot measurements without network pre-training, thereby improving the applicability of the HSIs reconstruction network.


\section{Coded aperture snapshot spectral imaging}

\begin{figure*}
\centering
\includegraphics [scale=0.51]{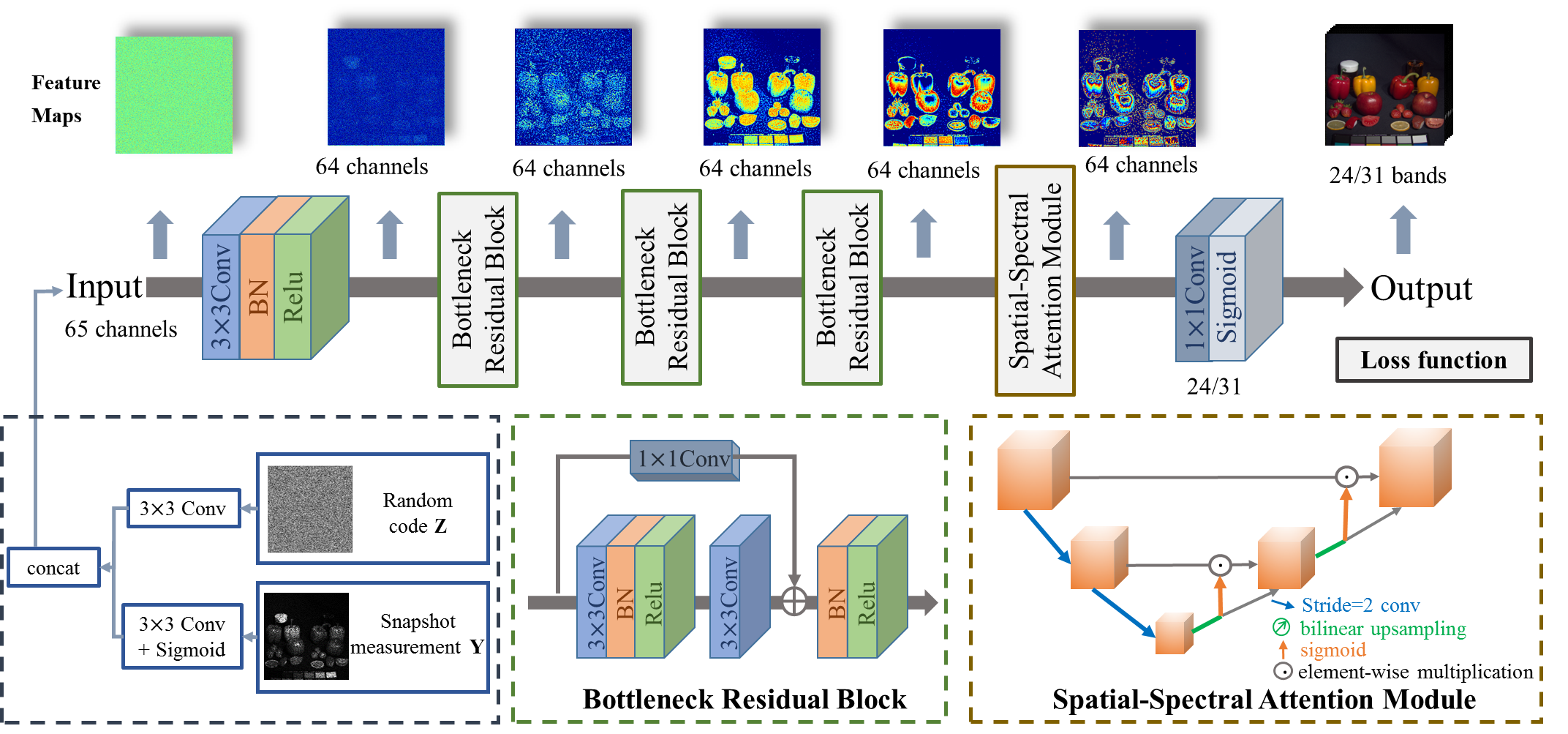}
\caption{The overall framework of the proposed HCS$^2$-Net. The snapshot measurement and the random code are taken as the inputs of HCS$^2$-Net, and multiple bottleneck residual blocks and one spatial-spectral attention module are exploited to reconstruct the original hyperspectral data-cube under the guidance of the loss function.}
\label{csnet}
\end{figure*}

The CASSI system makes full use of a coded aperture (physical mask) and one or more dispersive elements to modulate the optical field of a target scene, and achieves the projection from the 3D HSI data-cube into a 2D detector according to the specific sensing process. According to different methods of encoding spectral signatures, CASSI can be mainly divided into two categories: SD-CASSI \cite{wagadarikar2008single} using a single disperser encoded in the spatial domain and SS-CASSI \cite{lin2014spatial} or DD-CASSI \cite{gehm2007single} encoded in both spatial and spectral domains.


For concreteness, let $y(h,w,\lambda )$ indicate the discrete values of source spectral intensity with wavelength $\lambda $ at location $(h,w)$. A coded aperture mask creates coded patterns by its transmission function $T(h,w)$, while a dispersive prism produces a shear along one spatial axis based on a wavelength-dependent dispersive function $\psi (\lambda )$. Here, $\psi (\lambda )$ is assumed to be linear.

For spatial encoding, the imaging systems, such as SD-CASSI, first create a coding of the incident light field and then shear the coded field through a dispersive element. The final snapshot measurement at the 2D detector array can be represented as an integral over the spectral wavelength $\lambda $,
\begin{equation}\label{cassiy1}
  y(h,w) = \sum\limits_{\lambda  = 1}^C {T(h + \psi (\lambda ),w)X(h + } \psi (\lambda ),w,\lambda ).
\end{equation}

For spatial-spectral encoding, the imaging systems, such as DD-CASSI and SS-CASSI, have two dispersive elements, and a coded aperture is placed between them. Specifically, the imaging system disperses incident light field, and creates a coded field through the coded aperture mask, and employs additional optics to unshear this coding. The final snapshot measurement can be presented as
\begin{equation}\label{cassiy2}
  y(h,w) = \sum\limits_{\lambda  = 1}^C {T(h - \psi (\lambda ),w)X(h,w,\lambda )}.
\end{equation}
In summary, the CASSI imaging process can be rewritten in the following standard form of an underdetermined system,
\begin{equation}\label{cassivector}
  y = \Phi x + e,
\end{equation}
where $x \in {R^{N}}$ is the vectorized representation of the underlying 3D HSI $X \in {R^{H \times W \times C}}$ with $H \times W$ as its spatial resolution, $C$ as its number of spectral bands and $N$ computed as $HWC$, $y \in {R^{M}}$ denotes the vectorized formulation of the corresponding 2D snapshot measurement, $\Phi \in {R^{M\times N}}$ is the forward measurement matrix, and $e$ is the measurement noise. For the SD-CASSI system, the sensed measurement has the dimension of $H\times (W+C-1)$, so $M$ is computed as $H(W+C-1)$. Since a second dispersive element in the DD-CASSI system can undo the dispersion caused by the first one, the sensed measurement of DD-CASSI system has the same spatial dimension as $X$, and $M$ is accordingly computed as $HW$. The measurement rate can be computed as ${1 \mathord{\left/  {\vphantom {1 C}} \right.  \kern-\nulldelimiterspace} C}$.

In order to demonstrate the intrinsic structure of $\Phi$, the snapshot projection produce can be externalized as the following according to Eq. (\ref{cassiy2}),
\begin{equation}\label{cassimatrix}
  Y = \sum\limits_{i = 1}^C {{X_i}} \odot {S_i} + {E_i},
\end{equation}
where $Y \in {R^{H \times W}}$ is the matrix representation of the sensed snapshot measurement, $\odot$ is the Hadamard (element-wise) product, ${X_i} \in {R^{H \times W}}$ is the $i$-th spectral band and ${S_i}$ is the shifted coded mask corresponding to the $i$-th band, ${E_i} \in {R^{H \times W}}$ is the measurement noise. Specifically, for each pixel with a $C$-dimensional spectral vector, it is collapsed to form one pixel in the snapshot measurement. Thus, the measurement matrix $\Phi$ in Eq. (\ref{cassivector}) can be specialized as
\begin{equation}\label{cassimatrix2}
  \Phi  = [{\Phi _1},{\Phi _2},...,{\Phi _C}],
\end{equation}
where $\Phi_i \in {R^{HW\times HW}}$ is the diagonal matrix by taking the vectorized ${S_i}$ as its diagonal entries in the case of the CASSI system with spatial-spectral encoding. Under the framework of compressed sensing theory, \cite{jalali2019snapshot} studies theoretical analysis of snapshot CS systems and demonstrates that the reconstruction error of the CASSI system is bounded.

\section{Hyperspectral Compressive Snapshot Reconstruction}

This paper aims at learning the reconstruction only based on the 2D CASSI measurement with an unsupervised generative network. To reach this purpose, two issues need to be solved. One is how to construct the generative network of the unknown hyperspectral image, and the other is how to effectively estimate the network parameters based on the given snapshot measurement. In the following subsections, we will discuss the details.

\subsection{ Spatial-Spectral Reconstruction Network}

Fig. \ref{csnet} illustrates the architecture of the proposed HCS$^2$-Net for hyperspectral compressive snapshot reconstruction. For the purpose of making the generative network CS task-aware, the generative network is required to be conditional on the snapshot measurement $Y$. Thus, we concatenate feature maps of the latent random codes $Z$ and snapshot measurement $Y$ as the network inputs. Then the network inputs are processed through multiple $1\times 1$ bottleneck residual blocks and one spatial-spectral attention module, which are dedicated to capturing spatial-spectral correlation in hyperspectral images. Finally, a $1\times 1$ convolution is employed to adjust the number of the final output channels to be the same as the number of hyperspectral bands (for example, 24 bands and 31 bands). The sigmoid activation limits the output range to $[0, 1]$.

\textbf{Bottleneck Residual Block}. Residual blocks have been shown to perform well in image feature representation. Taking into account the close similarity between spectral bands, we specialize the residual blocks with $1\times1$ bottleneck connection and cascade three $1\times 1$ bottleneck residual blocks (dubbed as BRB) for feature extraction. Taking ${I^0}$ as the input of first residual block, the skip $1\times 1$ connection here is to fuse the correlation between bands as ${F_{spectral}}({I^0})$, and the main path can capture the remaining information ${F_{residual}}({I^0})$. Thus, the output ${I^1}$ of the first block is computed as
\begin{equation}\label{bottleneck}
  {I^1} = F_{BR}^1({I^0}) = {F_{residual}}({I^0}) + {F_{spectral}}({I^0}),
\end{equation}
where $F_{BR}^1$ denotes the operation in the first block. The subsequent two blocks further extract features from the previous block output, so the output ${I^3}$ of the $3$-th block can be expressed as,
\begin{equation}\label{bottleneckcascade}
  {I^3} = F_{BR}^3F_{BR}^2(F_{BR}^1({I^0})).
\end{equation}
This result is fed into the spatial-spectral attention module for further processing.

\textbf{Spatial-Spectral Attention Module}. Spatial-spectral joint correlation is an inherent characteristic of hyperspectral data-cube. At the same time, hyperspectral data also has multi-scales structure, just like in gray and color images. Thus, we design the spatial-spectral attention module to predict three-dimensional (3D) attention maps at multi-scales, so that these characteristics can be exploited to represent HSIs more effectively.

As shown in Fig. \ref{csnet}, the spatial-spectral attention module is conducted on multi-scale features and the 3D attention prediction is performed at each scale. The input of spatial-spectral attention module is $I^3$. We omit the superscript in this subsection for simplicity. Let ${I_i}$ represent the feature maps at scale $i$ $(i = 1,2,3)$, ${I_{i+1}}$ is computed from ${I_{i}}$ through a downsampling operation and $ 3 \times 3$ convolution, $i.e.$,
\begin{equation}\label{scaledown}
  {I_{i + 1}} = conv({I_i} \downarrow 2 ), i = 1,2,
\end{equation}
where $({I_i} \downarrow 2)$ is the downsampled feature maps from $I_{i + 1}$ by a convolution with stride 2. The spatial resolution of ${I_i}$ is $\frac{W}{{{2^{i - 1}}}} \times \frac{H}{{{2^{i - 1}}}}$, where $W,H$ is the spatial resolution of the feature maps $I_1$. Then, the feature maps at the $(i+1)$-th scale are used to compute the attention map to enhance the feature maps in the $i$-th scale. The computation flow is defined as,
\begin{equation}\label{scaleup}
\left\{ {\begin{array}{*{20}{c}}
{{A_i} = \sigma (conv({I_{i + 1}} \uparrow 2)),i = 1,2},\\
{{{I_i}^\prime} = {A_i} \odot conv({I_i}),{\rm{  }}i = 1,2},
\end{array}} \right.
\end{equation}
where $({I_{i + 1}} \uparrow 2 )$ denotes the double upsampled feature maps from $I_{i + 1}$ by bilinear interpolation, $A_i$ is the three-dimensional attention map for the $i$-th scale feature maps, $\odot$ denotes the Hadamard product. Specifically, we predict $A_i$ through the convolution and Sigmoid activation processing of $({I_{i + 1}} \uparrow2 )$. Different from the two-dimensional attention map or the tensor product of a two-dimensional attention and an one-dimensional attention map, we directly learn the 3D attention map, so that each entry of the feature maps can be adaptively weighted. Correspondingly, we obtain the attention enhanced feature maps ${{I_i}^\prime}$ by the Hadamard product operation between $A_i$ and $ conv({I_i})$. Furthermore, we concatenate ${{I_i}^\prime}$ with $({I_{i + 1}} \uparrow 2)$ to better fuse the spatial and spectral information among different scales. The final output at $i$-th scale are computed by the below formula,
\begin{equation}\label{scaleattention}
{\hat I_i} = conv(concat(I'_i,{I_{i + 1}} \uparrow 2)), i=1,2.
\end{equation}
$\hat I_1$ is then fed into the succeeding operations.

\subsection{Unsupervised Network Learning}


We define the conditional generative network in Fig. \ref{framework} as ${G_{{Z \mathord{\left/  {\vphantom {Z Y}} \right.  \kern-\nulldelimiterspace} Y}}}(w)$, in which $w$ is the network parameter, $Z $ is the random code and $Y $ is the snapshot measurement. The generative network ${G_{{Z \mathord{\left/
 {\vphantom {Z Y}} \right.  \kern-\nulldelimiterspace} Y}}}(w)$ acts as the parametric mapping from the latent random code $Z$ to the reconstruction conditioned on the snapshot
 measurement $Y$. With the aim of unsupervised learning, we try to learn the reconstruction only from the given snapshot measurement $Y$. According to the observation model in Eq. (\ref{cassimatrix}), the optimal reconstruction can be derived by solving the following loss function,

 \begin{equation}\label{loss}
   {w^*} = \arg \mathop {\min }\limits_w {\left\| {Y - \sum\limits_{i = 1}^C {{P_i}({G_{Z|Y}}(w)) \odot {S_i}} } \right\|_1},
 \end{equation}
where $P_i$ is the operation to extract the $i$-th band of the network output. Compared with the $l_2$ norm, the $l_1$ norm is more robust to outliers in the snapshot measurement \cite{li2004lae}. As in \cite{ulyanov2018deep}, we can seek the optimal $w$ so that the generated hyperspectral image can match the given measurement $Y$ as closely as possible. For different snapshot imaging systems, we only need to use the corresponding observation model in the loss function Eq. \ref{loss}. Therefore, the proposed network can well adapt to different imaging settings, which inherently enhances the applicability of our network in practical applications.


According to the chain rule of calculus, the approximation error can be backpropagated from the loss function $L$ to the network parameters $w$. The Adam \cite{kingma2014adam} algorithm is then used to find the optimal $w^*$ for reconstruction. This optimization is tailored to solve the reconstruction task for one specific image, and the final reconstruction can be computed from the optimal parameters $w^*$ as ${G_{{Z \mathord{\left/
 {\vphantom {Z Y}} \right.
 \kern-\nulldelimiterspace} Y}}}({w^ * })$.

\begin{figure}[htbp]
\centering
\includegraphics [scale=0.36]{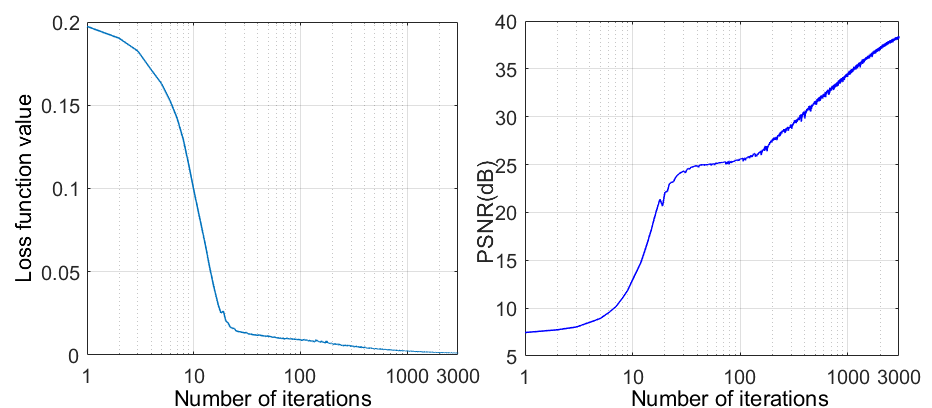}
\caption{A sample result of the Toy image from CAVE dataset. The left is the curve of loss function value versus the number of iterations, and the right is the curve of the PSNR value versus the number of iterations.}
\label{loss curves}
\end{figure}

Fig. \ref{loss curves} plots the variation of the loss function and PSNR versus the number of iterations in the optimization learning process. The number of iterations is in log-scale. This figure takes the Toy image from the CAVE dataset \cite{cave} as an example. It can be seen that during the iterative process, the value of the objective function first drops rapidly and finally stabilizes. As the value of the objective function decreases, the PSNR values of the reconstruction results generally show an upward trend, although there are some fluctuations, and finally we can obtain a good reconstruction result.

\section{Experimental Results}
We evaluate the reconstruction performance of the proposed HCS$^2$-Net and compare it with multiple state-of-art methods, including predefined prior-based-based methods, $i.e.$, TwIST \cite{bioucas2007new}, GAP-TV \cite{yuan2016generalized}, DeSCI \cite{liu2018rank}, and deep network-based methods, $i.e.$, AutoEncoder \cite{choi2017high}, HSCNN \cite{xiong2017hscnn}, HyperReconNet \cite{wang2018hyperreconnet}, $\lambda$-net \cite{miao2019lambda} and the residual network (dubbed as DEIL) in \cite{zhang2019hyperspectral}. Same as the above methods, two typical image quality metrics are used to evaluate the performance, namely peak-signal-to-noise-ratio (PSNR) and structural similarity index (SSIM \cite{wang2004ssim}). PSNR can reflect the spectral reflectance accuracy and SSIM emphasizes spatial reflectance accuracy of the reconstruction. The larger the PSNR and SSIM values are, the better the reconstruction accuracy is.

For a comprehensive evaluation, the reconstruction experiments are conducted upon both simulated and real CASSI measurements, and both the SS-CASSI and SD-CASSI systems are tested. The simulated CASSI measurements include two cases, namely synthetic coded aperture masks and the “Real-Mask-in-the-Loop” masks from the real CASSI systems. The simulation experiments are conducted on two datasets, CAVE \cite{cave} and ICVL \cite{ICVL}. The CAVE dataset contains 32 scenes with spatial size $512\times512$ and the number of spectral bands 31. The range of wavelengths of CAVE covers from 400nm to 700nm with uniform 10nm intervals. The spatial resolution of images in the ICVL dataset is $1392\times1300$, and the number of spectral bands is 31. As in \cite{miao2019lambda}, the spectral resolution of ICVL dataset are downsampled to 24 bands. Besides the CAVE and ICVL datasets, we also test the proposed HCS$^{2}$-Net on the real $Bird$ and $Leaves$ hyperspectral data.

Autoencoder \cite{choi2017high}, HSCNN \cite{xiong2017hscnn}, HyperReconNet \cite{wang2018hyperreconnet} and DEIL \cite{zhang2019hyperspectral} are four supervised networks, and they select some HSIs from three datasets including CAVE, Harvard \cite{ayan2011harvard} and ICVL for network pre-training. $\lambda$-net \cite{miao2019lambda} is pre-trained for reconstructing hyperspectral data with 24 spectral bands, and 150 hyperspectral data after spectral interpolation processing are selected from ICVL for network training. Different from the above four deep networks, the proposed HCS$^2$-Net is an unsupervised deep network and does not require network pre-training. We implement HCS$^{2}$-Net using the Pytorch framework and all the experiments are performed on an NVIDIA GTX 1080 Ti GPU. In our experiments, the random code $Z$ is initialized as random noise maps with uniform distribution, the learning rate is set to 0.01. The optimizer adopts the Adam method, and the maximum number of iterations is set to 2500, which can achieve a good balance between reconstruction accuracy and reconstruction time.

\newcommand{\tabincell}[2]{\begin{tabular}{@{}#1@{}}#2\end{tabular}}
\begin{table*}[htbp]
\renewcommand\arraystretch{1.5}
 \caption{Performance evaluation of the proposed  method under different ablation settings.}

 \resizebox{\textwidth}{!}{
 \begin{tabular}{|c|c|c|c|c|c|c|}
  \hline
   \multicolumn{2}{|c|}{Setting} &Ablation method 1	&Ablation method 2	&Ablation method 3	&Ablation method 4	&HCS$^2$-Net \\
  \hline
   \multirow{3}{*}{\tabincell{c}{Network Input \\ evaluation}} & only random code as input     & \checkmark	 & $\times$	  &$\times$    &$\times$    &$\times$ \\
                     \cline{2-7}             & only measurement as input                     & $\times$      &\checkmark  &$\times$    &$\times$    &$\times$  \\
                     \cline{2-7}             & random codes+ measurement as inputs                   & $\times$	     &$\times$	  &\checkmark  &\checkmark   &\checkmark  \\
   \hline
   \multirow{3}{*}{\tabincell{c}{Network Architecture \\evaluation}} & only residual block   & $\times$	     &$\times$	  &$\times$     &\checkmark	 &$\times$ \\
                     \cline{2-7}                    & only attention module                  & $\times$	     &$\times$	  &\checkmark   &$\times$	 &$\times$ \\
                     \cline{2-7}                    & residual block+ attention module       & \checkmark	 &\checkmark  &$\times$      &$\times$	 &\checkmark \\
   \hline
   \multirow{2}{*}{metrics}         & PSNR/SSIM (SD-CASSI)    &28.277/0.876   &28.751/0.909  & 28.304/0.890  &29.256/0.895 	 & 30.764/0.930 \\
                     \cline{2-7}    & PSNR/SSIM (SS-CASSI)    &34.629/0.948   &36.026/0.955  &35.607/0.950   &36.703/0.964	&39.219/0.979  \\
   \hline
 \end{tabular}}
\label{ablation}
\end{table*}

\subsection{Ablation studies}

We first conduct some ablation experiments to evaluate the influences of different settings of HCS$^2$-Net on the reconstruction results. These settings fall into three categories. One is related to the network inputs, including only the random code as the input, only the snapshot measurement as the input and both the random code and snapshot measurement as the inputs. The other is related to the network architecture, including the cases of only the 1$\times$1 residual blocks, only the spatial-spectral attention module, and the complete architecture ( as shown in Fig. \ref{csnet}). We choose appropriate settings to form four ablation methods, where ablation method 1 and 2 are used to verify the network inputs and ablation method 3 and 4 are used to verify the network architecture. Table \ref{ablation} reports the experimental results of four ablation methods and HCS$^2$-Net upon all the 32 scenes of CAVE dataset. We conduct the ablation experiments under both cases of SD-CASSI and SS-CASSI, and give the average PSNR and SSIM values of each ablation method and HCS$^2$-Net in these two cases.

According to the average PSNR and SSIM values of the ablation methods 1 and 2 and HCS$^2$-Net, we can see that taking both the snapshot measurement and random code as the inputs can obtain significant performance improvements in both cases of SD-CASSI and SS-CASSI. We analyze that this performance improvement mainly comes from two aspects. One is that using the snapshot measurement as a conditional input can make the generator aware of the reconstruction task, and the other is that the snapshot measurement contains the spatial structure of the underlying scene, so the convolutions on the snapshot measurement can still extract useful features for reconstruction. The comparison between the ablation methods 3 and 4 with HCS$^2$-Net is supposed to verify the contributions of the 1$\times$1 residual blocks and the spatial-spectral attention module for reconstruction. According to Table \ref{ablation}, it can be also seen that the absence of anyone module will result in performance degradation in both cases of SD-CASSI and SS-CASSI. The combination of the 1$\times$1 residual blocks and the spatial-spectral attention module can achieve the best reconstruction, thus the ablation experiments demonstrate that the design of HCS$^2$-Net is very reasonable.

\begin{table}[ht]
\begin{center}
 \caption{\label{SS-CASSI results} Average PSNR (dB) and SSIM performance comparisons of various methods on the whole CAVE dataset using the SS-CASSI measurement. The best performance is labeled in bold.}
 \begin{tabular}{llll}
  \toprule
  Methods & PSNR & SSIM  \\
  \midrule
  TwIST \cite{bioucas2007new}  & 23.74 & 0.85  \\
  GAP-TV \cite{yuan2016generalized} & 27.15 & 0.89  \\
  DeSCI \cite{liu2018rank} & 29.20 & 0.91  \\
  AutoEncoder \cite{choi2017high}  & 32.46 & 0.95  \\
  HCS$^{2}$-Net & \textbf{39.22} & \textbf{0.98} \\
  \bottomrule
 \end{tabular}
\end{center}
\end{table}

\begin{table}[ht]
\begin{center}
 \caption{\label{SD-CASSI results} Average PSNR (dB) and SSIM performance comparisons of various methods on the 10 scenes of CAVE dataset using the SD-CASSI measurement.}
 \begin{tabular}{lll}
  \toprule
  Methods & PSNR & SSIM  \\
  \midrule
  HSCNN \cite{xiong2017hscnn} & 25.02 & 0.91 \\
  AutoEncoder \cite{choi2017high} & 25.74 & 0.91 \\
  HyperReconNet \cite{wang2018hyperreconnet} & 24.44 & 0.90 \\
  DEIL \cite{zhang2019hyperspectral} & 29.05 & \textbf{0.95}\\
  HCS$^{2}$-Net & \textbf{29.33} & 0.928 \\
  \bottomrule
 \end{tabular}
\end{center}
\end{table}

\subsection{Results on Synthetic Measurement}

\begin{figure}
\centering
\includegraphics [scale=0.18]{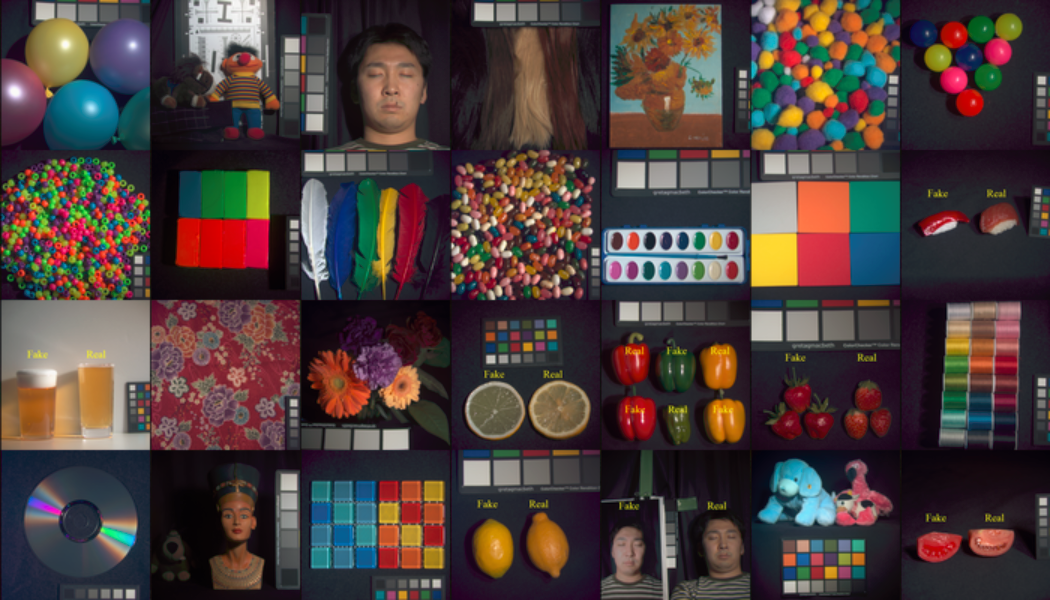}
\caption{Some representative scenes in CAVE dataset.}
\label{cave}
\end{figure}

We conduct experiments on the CAVE dataset under two synthetic measurement scenarios, $i.e.$, SS-CASSI and SD-CASSI. Fig. \ref{cave} displays the RGB images of some representative scenes in the CAVE dataset. The coded masks used here are randomly generated according to the corresponding SD-CASSI or SS-CASSI principles.


\begin{figure*}
\centering
\includegraphics [scale=0.31]{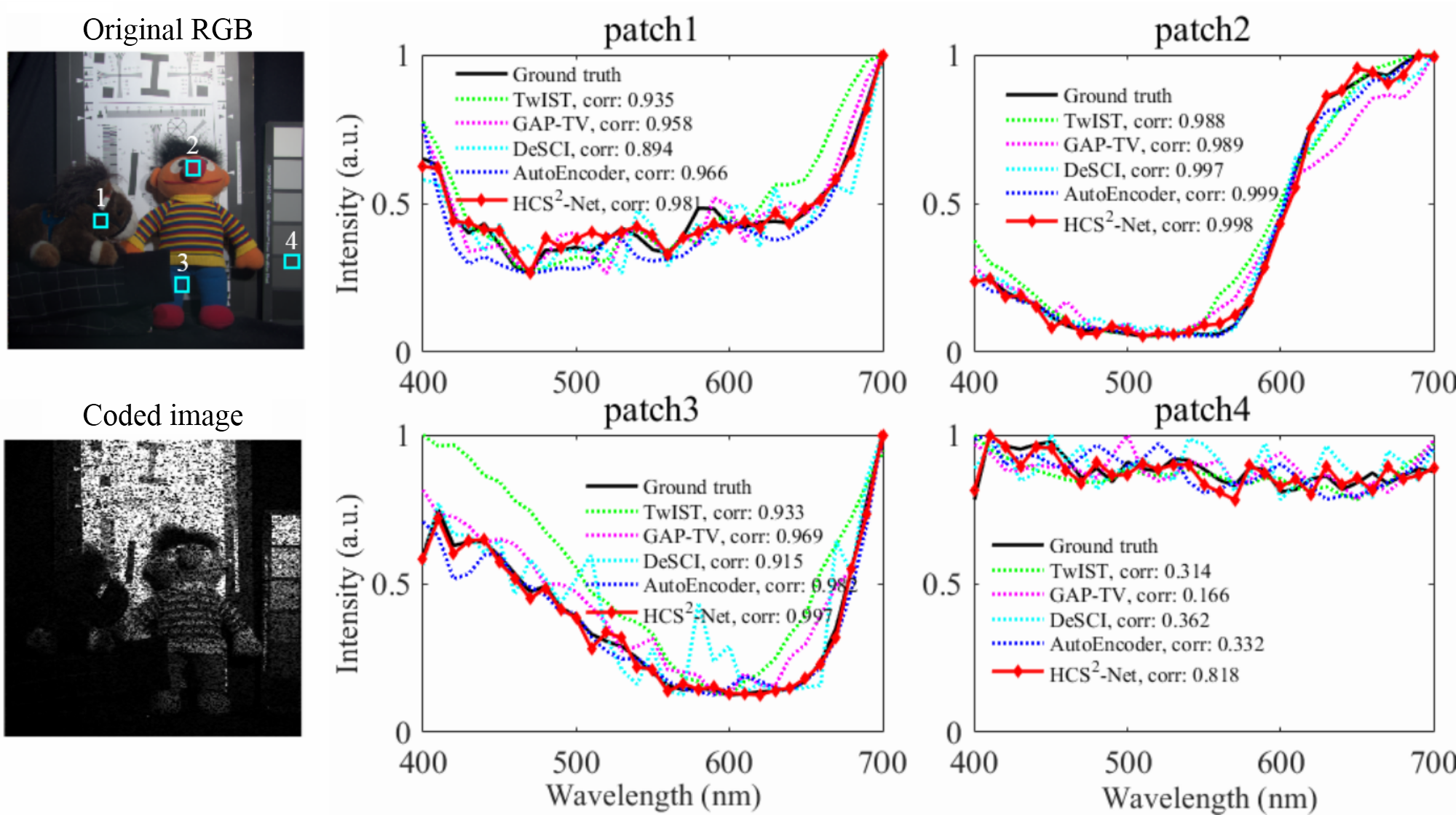}
\caption{The RGB image of the Toy scene from the CAVE dataset and the corresponding snapshot measurement by the SS-CASSI encoding. The Toy scene has the spatial and spectral resolution of $512\times512\times 31$. The right two columns show the reconstructed spectral signatures corresponding to the four patches indicated in the RGB image and the corresponding ground-truths. Correlation coefficients are also calculated to quantitatively evaluate the accuracy of the reconstructed spectral signatures of the five methods.}%
\label{toy image1}
\end{figure*}

\begin{figure*}
\centering
\includegraphics [scale=0.41]{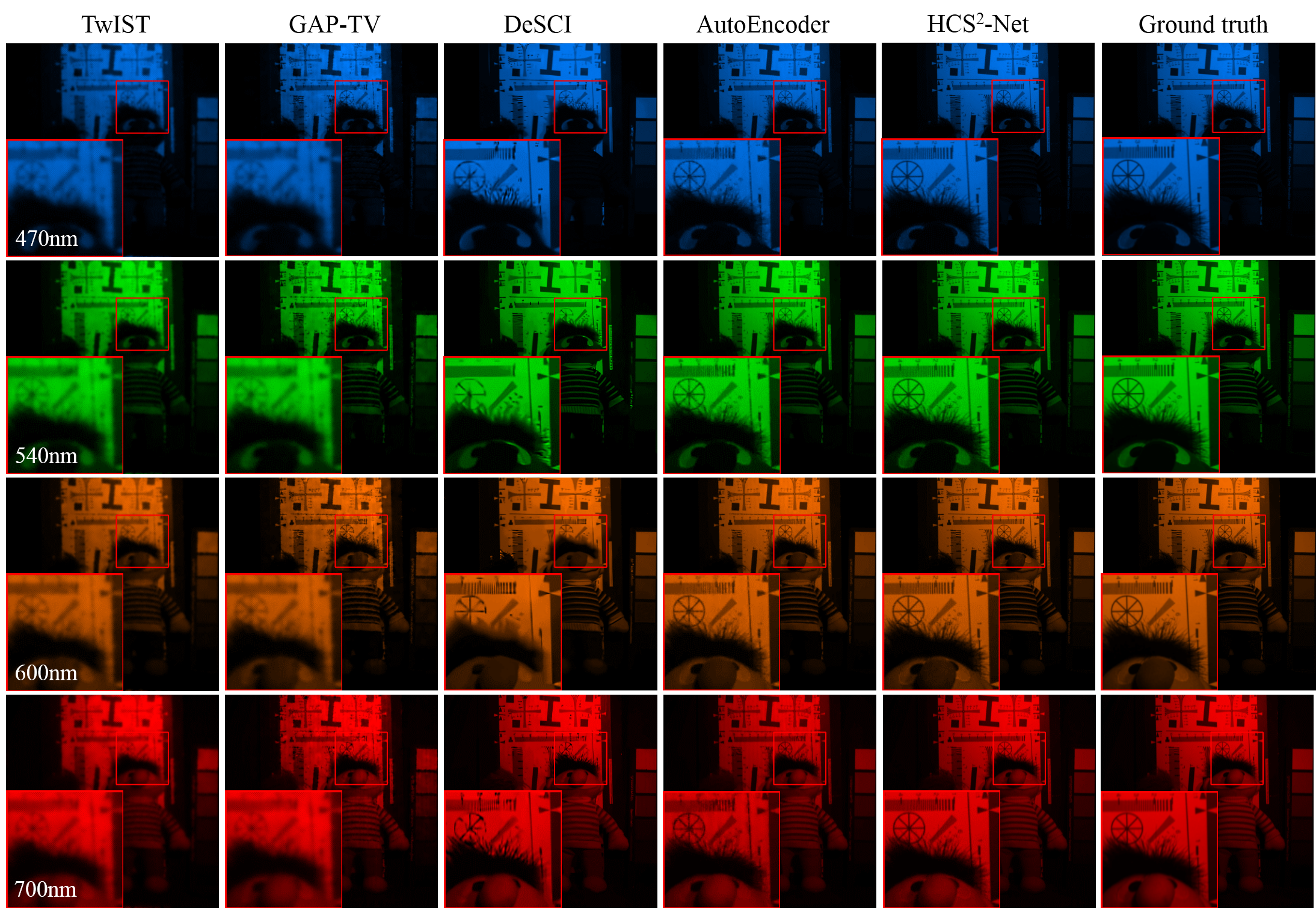}
\caption{Reconstructed spectral images of the Toy scene from the CAVE dataset. We visualize four spectral bands (wavelength: 470nm, 540nm, 600nm and 700nm) reconstructed by five methods, and the PSNR and SSIM values of the reconstructed Beads scene by each method are TwIST (25.87/0.84), GAP-TV (27.91/0.92), DeSCI (28.35/0.92), AutoEncoder (33.52/0.98) and HCS$^2$-Net (39.79/0.99). } 
\label{toy image2}
\end{figure*}

Table \ref{SS-CASSI results} presents the average PSNR (dB) and SSIM values of various methods upon all the 32 scenes in the CAVE dataset using SS-CASSI measurement.
HCS$^2$-Net has significant superiority over the predefined prior-based methods and pre-trained networks in terms of both PSNR (dB) and SSIM metrics. To visualize the experimental results, the reconstructed results of five algorithms on the Toy image are shown in Fig. \ref{toy image1} and  Fig. \ref{toy image2}. We use the wavelength-to-RGB converter to display each band of the spectral reconstruction results, and display four spectral bands. In order to clearly compare the details of the reconstructed images, We also provide zoomed-in views of some image areas represented by the rectangles. The reconstructed spectral signatures at four positions indicated in the RGB images are also presented. The correlation coefficients of the reconstructed spectral signatures and the Ground-truths are shown in the legends. By comparing the reconstructed spectral bands and spectral signatures with the ground-truths, it can be clearly seen that HCS$^2$-Net is superior to the other four comparison methods. Our method can effectively preserve the spatial structures and the spectral accuracy of the reconstructed hyperspectral image, thus demonstrating the capability of HCS$^2$-Net to utilize the inherent characteristics of hyperspectral images and verifying the effectiveness of unsupervised learning.

Table \ref{SD-CASSI results} lists the reconstruction results using the SD-CASSI measurement. As in \cite{xiong2017hscnn, wang2018hyperreconnet,zhang2019hyperspectral}, we adopt the same 10 scenes from the CAVE dataset as the test set for a fair comparison. The results of the other four methods in Table \ref{SD-CASSI results} are cited from \cite{zhang2019hyperspectral}. We can see that HCS$^2$-Net also achieves superior or comparable results over four state-of-the-art deep networks, although they use a large number of data sets for pre-training. The superior performance of HCS$^2$-Net mainly benefits from the conditional generative manner and the spatial-spectral attention module, which can more effectively capture the inherent spatial-spectral correlations of HSIs. In addition, the characteristics of unsupervised learning can also enhance the adaptability and universality of our network in practical applications.

\begin{figure}[htbp]
\centering
\includegraphics [scale=0.32]{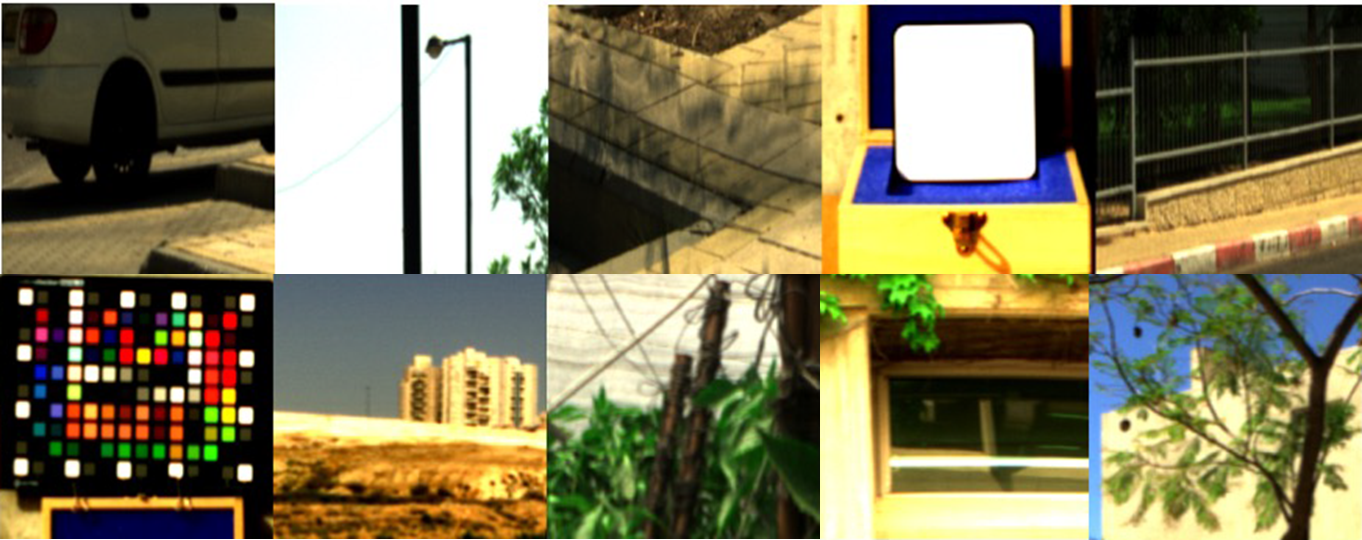}
\caption{10 testing scenes in the ICVL dataset.}
\label{10testing imgs}
\end{figure}

\begin{figure*}[htbp]
\centering
\includegraphics [scale=0.38]{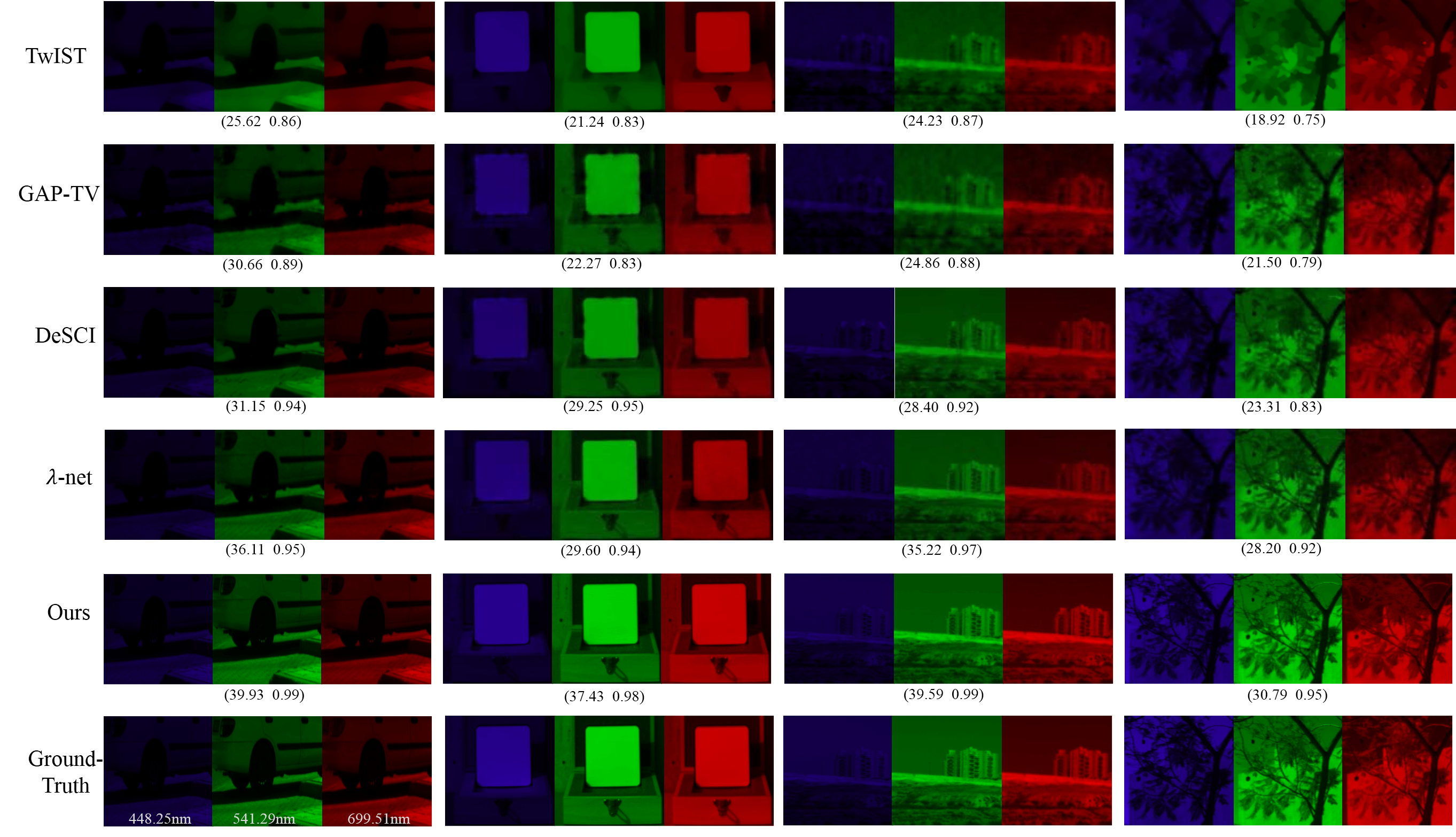}
\caption{Reconstructed spectral images of four scenes from the ICVL hyperspectral dataset. The three spectral bands (with wavelengths 448.25nm, 541.29nm and 699.51nm) of each reconstructed hyperspectral image are selected for visualization. The reconstructions by five methods (TwIST/GAP-TV/DeSCI/$\lambda$-net/HCS$^2$-Net) are shown from top to bottom, and the corresponding ground-truths are on the last row.}

\label{lambda results}
\end{figure*}

\begin{table*}
\renewcommand\arraystretch{1.45}
  \centering
  \resizebox{\textwidth}{!}{
\begin{tabular}{c|c|c|c|c|c|c|c|c|c|c|c|c}
  \hline
  \hline
  Methods              & Metrics & scene 1 & scene 2 & scene 3 & scene 4 & scene 5 & scene 6 & scene 7 & scene 8 & scene 9 & scene 10 & Average\\
  \hline
  \multirow{2}{*}{TwIST \cite{bioucas2007new}}  & PSNR & 25.62 & 18.41 & 21.75 & 21.24 & 23.78& 20.58 & 24.23 & 20.20& 27.01 & 18.92 & 22.14\\
         \cline{2-13}     & SSIM & 0.856 & 0.826 & 0.826 & 0.828 & 0.799 & 0.744 & 0.870 & 0.784 & 0.888 & 0.747 & 0.817\\
  \hline
  \multirow{2}{*}{GAP-TV \cite{yuan2016generalized}} & PSNR & 30.66 & 22.41 & 23.49 & 22.27 & 26.98 & 23.09 & 24.86 & 22.91 & 29.10 & 21.50 & 24.73\\
          \cline{2-13}    & SSIM & 0.892 & 0.869 & 0.863 & 0.829 & 0.792 & 0.802 & 0.877 & 0.841 & 0.912 & 0.796 & 0.847\\
  \hline
  \multirow{2}{*}{DeSCI \cite{liu2018rank} }  & PSNR & 31.15 & 26.44 & 24.74 & 29.25 & 29.37 & 25.81 & 28.40 & 24.42 & 34.41 & 23.31 & 27.73\\
           \cline{2-13}   & SSIM & 0.937 & 0.947 & 0.898 & 0.949 & 0.907 & 0.906 & 0.921 & 0.872 & 0.971 & 0.834 & 0.914\\
  \hline
  \multirow{2}{*}{$\lambda$-Net \cite{miao2019lambda}} & PSNR & 36.11 & 32.05 & 33.34 & 29.60 & \textbf{35.40} & \textbf{28.57} & 35.22 & 32.35 & \textbf{33.42} & 28.20 & 32.43\\
           \cline{2-13}   & SSIM & 0.949 & 0.975 & 0.974 & 0.937& 0.942 &\textbf{0.902} & 0.969 & 0.951 & 0.916 & 0.924 & 0.944\\
  \hline
  \multirow{2}{*}{HCS$^2$-Net} & PSNR & \textbf{39.94} & \textbf{36.74} & \textbf{36.30} &\textbf{ 37.43} & 32.07 & 24.06 & \textbf{39.59} & \textbf{35.70} & 32.57 & \textbf{30.79} & \textbf{34.52}\\
           \cline{2-13}   & SSIM & \textbf{0.990} & \textbf{0.992} & \textbf{0.981} & \textbf{0.984} & \textbf{0.963} & 0.899 & \textbf{0.991} &\textbf{ 0.978} & \textbf{0.974} & \textbf{0.955} & \textbf{0.970} \\
  \hline
  \hline
\end{tabular}}
\caption{Average PSNR (dB) and SSIM values of five methods upon 10 scenes from the ICVL dataset by using the “Real-Mask-in-the-Loop” coded masks.}
\label{lambda10 results}
\end{table*}

\subsection{Results on “Real-Mask-in-the-Loop”}

We further test the performance of different methods using the “Real-Mask-in-the-Loop” coded masks, that is, the masks used here is from the real CASSI systems \cite{miao2019lambda}. Compared to the mask generated by simulation, real masks contain more noise, which makes the reconstruction more difficult. For the sake of fairness, we adopt the same 10 HSIs of the ICVL dataset used in $\lambda$-net \cite{miao2019lambda} for comparison. The RGB images of these 10 HSIs are shown in Fig. \ref{10testing imgs}. As in $\lambda$-net \cite{miao2019lambda}, we perform spectral interpolation and cropping operations upon these 10 HSIs, and their spectral and spatial resolutions become 24$\times$256$\times$256. The SS-CASSI measurement is used in this group of experiments.

Table \ref{lambda10 results} lists the PSNR and SSIM values and their average values of the reconstruction results of multiple methods on the 10 scenes from the ICVL dataset. Because HSCNN, AutoEncoder, HyperReconNet and DEIL are trained for reconstructing HSIs of 31 spectral channels, thus they are not applicable to this group of experiments. Fig. \ref{lambda results} shows three spectral bands in the reconstructed hyperspectral images of five algorithms. According to the results in Table \ref{lambda10 results}, HCS$^2$-Net exceeds both the predefined prior-based reconstruction algorithms and the deep network-based reconstruction algorithms, and it obtains the best reconstruction quality in terms of the average PSNR and SSIM values. From the reconstructed spectral images of the 4 scenes in Fig. \ref{lambda results}, we can see that HCS$^2$-Net can reconstruct more clear structures and details than the competing methods.

\subsection{Results on Real Compressive Snapshot Imaging Data}

\begin{figure*}
\centering
\includegraphics [scale=0.20]{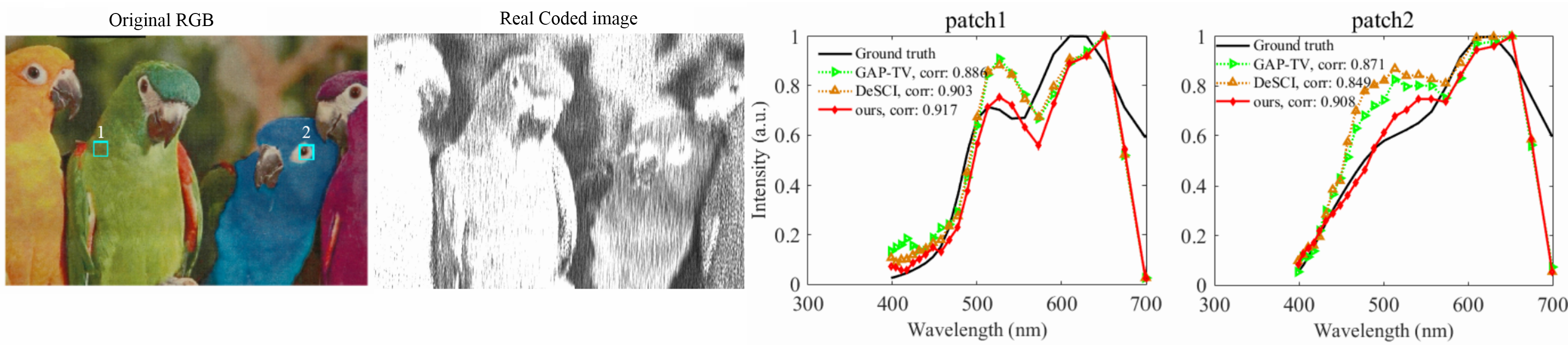}
\caption{Real data results: the reconstructed spectral signatures of the real $Bird$ hyperspectral data captured by the real CASSI system. The correlation coefficient of the reconstructed spectral and the ground-truth is shown in the legends.}
\label{realdataimage1}
\end{figure*}

\begin{figure*}
\centering
\includegraphics [scale=0.40]{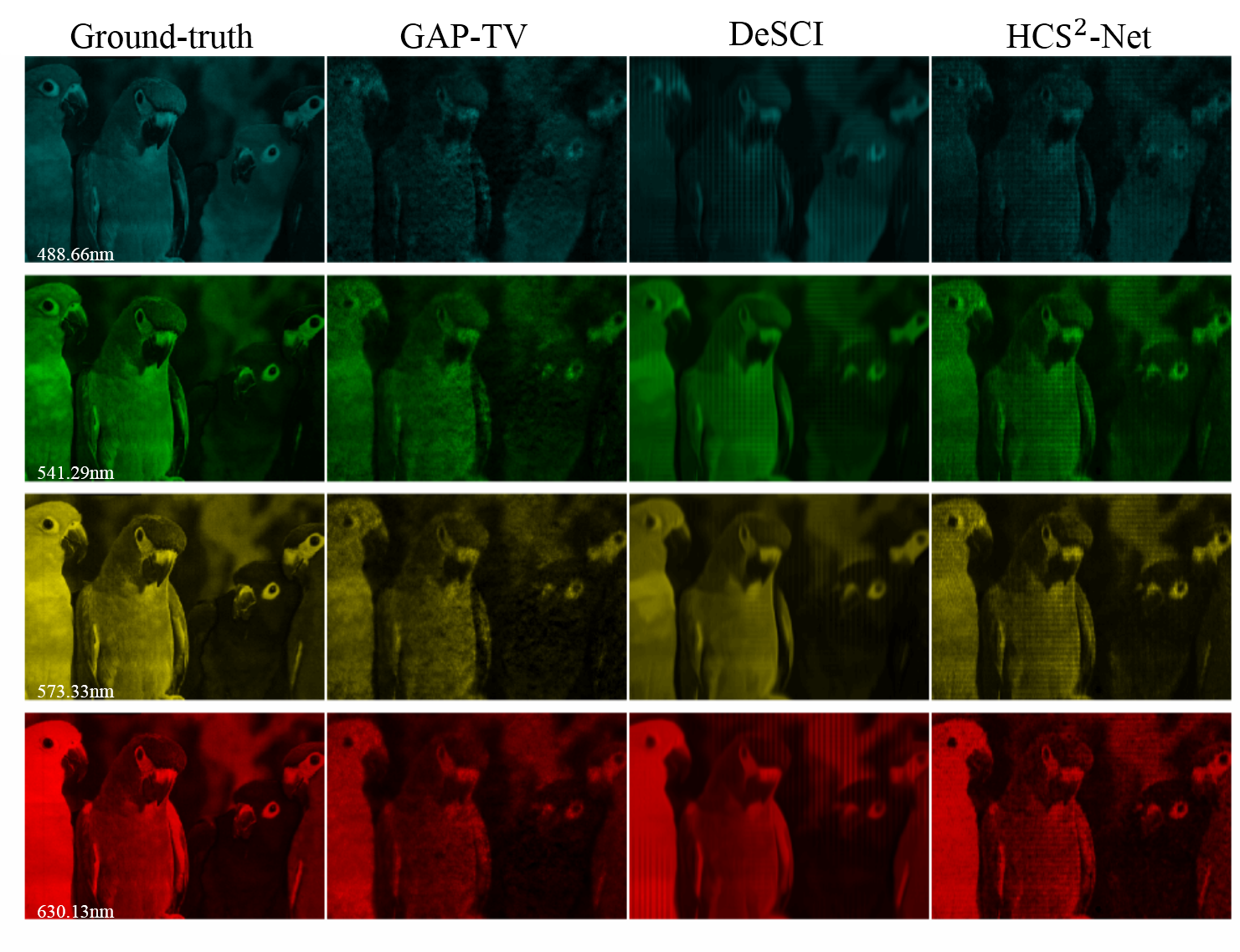}
\caption{Real data results: Four spectral bands (wavelength: 488.67nm, 541.29nm, 573.33nm and 630.13nm) of the real $Bird$ hyperspectral data reconstructed by three methods, and the corresponding PSNR and SSIM values of each method are GAP-TV (21.99/0.67), DeSCI (24.50/0.67) and HCS$^2$-Net (24.69/0.70).}
\label{realdataimage2}
\end{figure*}


In order to demonstrate the performance of HCS$^2$-Net more persuasively, we further perform the experiments on the real hyperspectral compressive snapshot imaging data, $i.e.$, the $Bird$ hyperspectral image$\footnote{The bird hyperspectral image is downloaded from \cite{liu2018rank}'s Github homepage https://github.com/hust512/DeSCI.}$. $Bird$ hyperspectral image is captured by the real CASSI system \cite{gehm2007single}. $Bird$ consists of 24 spectral bands with the spatial resolution $1021\times703$. The wavelengths of 24 spectral bands of $Bird$ data are: {398.62, 404.40, 410.57, 417.16, 424.19, 431.69, 439.70, 448.25, 457.38, 467.13, 477.54, 488.66, 500.54, 513.24, 526.80, 541.29, 556.78, 573.33, 591.02, 609.93, 630.13, 651.74, 674.83, 699.51} nm. 
The real imaging system means a more daunting challenge, as the 2D CASSI coded images captured by the real snapshot compressive imaging system are companied by more noise and outliers.

The reconstructed spectral signatures and exemplary bands of the $Bird$ image are shown in Fig. \ref{realdataimage1} and Fig. \ref{realdataimage2}. We also plot two reconstructed spectral signatures corresponding to the two positions indicated in the RGB images. HCS$^2$-Net has a superior performance over GAP-TV and DeSCI in terms of the PSNR and SSIM values. As shown in the reconstructed spectral bands, the reconstruction of GAP-TV still contains noise, and DeSCI produces excessively smooth results, leading to the loss of some details. In contrast, HCS$^2$-Net can recover detailed structures, resulting in relatively good reconstruction quality. Moreover, HCS$^2$-Net can reconstruct more accurate spectral signatures than DeSCI. Accurate reconstruction of spectral signatures is important for applications such as material deification and classification. It implies that HCS$^2$-Net has the potential to promote the development of hyperspectral compressive snapshot imaging technology.


\subsection{Time Complexity}

\begin{table}[ht]
\begin{center}
 \caption{\label{time_results}The running-time (in seconds) of various methods to reconstruct a 512$\times$512 HSI of the CAVE dataset.}
 \begin{tabular}{cccccc}
  \toprule
  Methods & \makecell*[c] {TwIST \\\cite{bioucas2007new}}  & \makecell*[c]{GAP-TV \\\cite{yuan2016generalized}} & \makecell*[c] {DeSCI \\ \cite{liu2018rank}} & \makecell*[c]{Auto \\Encoder\\ \cite{choi2017high}}  & \makecell*[c]{HCS$^2$\\-Net}\\
  \midrule
  Time (s) & 441 & 49 & 14351 & 414 & 367 \\
  \bottomrule
 \end{tabular}
\end{center}
\end{table}

We further analyze the time complexity of the proposed method and other baselines for hyperspectral compressive snapshot reconstruction. Table \ref{time_results} shows the running time required for each method to reconstruct a 512$\times$512 HSI of the CAVE dataset. We use the source codes provided by the authors on the web to run the comparison algorithms including TwIST, GAP-TV, DeSCI and Autoencoder. The source codes of TwIST, GAP-TV and DeSCI are written by matlab and run on the CPU. AutoEncoder and the proposed HCS$^2$-Net are written by deep learning frameworks and run on the GPU. According to Table \ref{time_results}, GAP-TV is relatively fast, but the reconstruction performance of this algorithm is far from satisfactory. DeSCI has the highest computational complexity, which is mainly due to the time-consuming operations of block matching and weighted nuclear norm minimization in each iteration of optimization. Compared with DeSCI, the running time of HCS$^2$-Net is obviously much shorter. The running time of AutoEncoder is comparable to HCS$^2$-Net, but the reconstruction quality is worse than our algorithm. In addition, the AutoEncoder method requires training time to learn the spectral prior. The proposed HCS$^2$-Net does not require pre-training on a large amount of hyperspectral data, so it will not consume training time. Although HCS$^2$-Net needs iterative optimization in the reconstruction, it maintains an acceptable time complexity of network learning on the GPU. At the same time, the unsupervised learning feature of HCS$^2$-Net can make it more adaptable for different hyperspectral imaging systems.




\section{Conclusion}
In this paper, we proposed the unsupervised HCS$^2$-Net for compressive snapshot reconstruction of HSIs. The proposed HCS$^2$-Net serves as a parametric mapping from the combination of the latent random code and snapshot measurement to the reconstruction. The $1\times1$ bottleneck residual block and spatial-spectral attention module can boost the network to capture the inherent spatial spectral correlation, thereby generating spatial-spectral structures more precisely. Furthermore, HCS$^2$-Net is optimized to generate the reconstruction from the given snapshot measurement, and this is fully unsupervised, requires no training data. HCS$^2$-Net is applicable to both the SD-CASSI and SS-CASSI systems. According to the experimental results, it is striking that HCS$^2$-Net can achieve superior or comparable reconstruction results over the state-of-the-art methods, including the deep networks with pre-training.

\bibliographystyle{IEEEtran}
\bibliography{IEEEabrv,reference}

\begin{IEEEbiography}[{\includegraphics[width=1in,height=1.25in,clip,keepaspectratio]{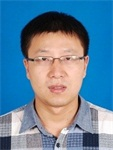}}]{Yubao Sun}
received the Ph.D. degree from Nanjing University of Science and Technology, Nanjing, China, in 2010. He is an associate professor in the School of Computer and Software, Nanjing University of Information Science and Technology, China. His research interests are deep learned based compressed sensing, low rank and sparse representation, video analysis, hypergraph learning, and so on.
\end{IEEEbiography}

\begin{IEEEbiography}[{\includegraphics[width=1in,height=1.25in,clip,keepaspectratio]{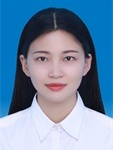}}]{Ying Yang}
received the bachelor’s degree in automation from the Nanjing University of Information Science and Technology, Nanjing, China, in 2018. She is current a master with a focus on image processing and compressed sensing reconstruction.
\end{IEEEbiography}

\begin{IEEEbiography}[{\includegraphics[width=1in,height=1.25in,clip,keepaspectratio]{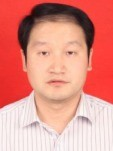}}]{Qingshan Liu}
received the M.S. degree from Southeast University, Nanjing, China, in 2000, and the Ph.D. degree from the Chinese Academy of Sciences, Beijing, China, in 2003. He was an Associate Professor with the National Laboratory of Pattern Recognition, Chinese Academy of Sciences. From 2010 to 2011, he was an Assistant Research Professor with the Department of Computer Science, Computational Biomedicine Imaging and Modeling Center, Rutgers University, The State University of New Jersey, Piscataway, NJ, USA. From 2004 to 2005, he was an Associate Researcher with the Multimedia Laboratory, The Chinese University of Hong Kong, Hong Kong. He is currently a Professor with the School of Computer and Software, Nanjing University of Information Science and Technology, Nanjing. His research interests include image and vision analysis.

Dr. Liu received the National Science Fund for Distinguished Young Scholars in 2018 and the President Scholarship of the Chinese Academy of Sciences in 2003.
\end{IEEEbiography}

\begin{IEEEbiography}[{\includegraphics[width=1in,height=1.25in,clip,keepaspectratio]{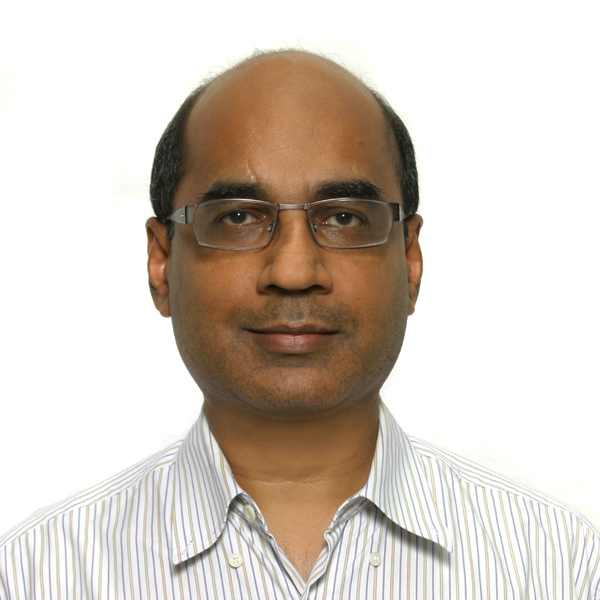}}]{Mohan Kankanhalli}
is the Provost’s Chair Professor at the Department of Computer Science of the National University of Singapore. He is the Director of N-CRiPT and also the Dean, School of Computing at NUS. Mohan obtained his BTech from IIT Kharagpur and MS \& PhD from the Rensselaer Polytechnic Institute. His current research interests are in Multimedia Computing, Multimedia Security \& Privacy, Image and Video Processing and Social Media Analysis. He is active in the Multimedia Research Community and is on the editorial boards of several journals. Mohan is a Fellow of IEEE.
\end{IEEEbiography}

\end{document}